# *In Vitro* Studies of Cells Grown on the Superconductor $PrO_xFeAs$

Shaoguang Yang[1*], Yuxuan Xie[2], Wenrong Yang[2], Rongkun Zheng[2], Frankie Stevens[2], Emine Korkmaz[2], Anthony S. Weiss[3], Simon P. Ringer[2], Filip Braet[2*]

* Corresponding authors

sgyang@nju.edu.cn   and   f.braet@usyd.edu.au

1 National Laboratory of Microstructures, Nanjing University, Nanjing, China

2 Australian Key Centre for Microscopy and Microanalysis, The University of Sydney

3 School of Molecular and Microbial Biosciences, The University of Sydney, Sydney, Australia.

Abstract

The recent discovery of arsenic-based high temperature superconductors has reignited interest in the study of superconductor : biological interfaces. However, the new superconductor materials involve the chemistry of arsenic, their toxicity remain unclear [ Nature, 2008, 452(24):922]. In this study the possible adverse effects of this new family of superconductors on cells have been examined. Cell culture studies in conjunction with microscopy and viability assays were employed to examine the influence of arsenic-based superconductor $PrO_xFeAs$ (x=0.75) material *in vitro*. Imaging data revealed that cells were well adhered and spread on the surface of the superconductor. Furthermore, cytotoxicity studies showed that cells were unaffected during the time-course of the experiments, providing support for the biocompatibility aspects of $PrO_xFeAs$-based superconductor material.

1. Introduction

There is substantial scientific interest in the recently identified FeAs based superconductors [1] and the development of high temperature superconductors using solid state materials. Ramped up efforts include element substitution with the goal of improving their superconductive properties [2-6]. Similar to the first generation of copper-based high temperature superconductor, the new superconductor has a sandwich structure of alternate conducting and insulating layers, where the superconducting current occurs in the FeAs layers. These second generation materials have broken the tyranny of copper based high temperature superconductors by showing that high-temperature conductivity is not the sole preserve of copper oxides. This concept has been expanded by observations where both electron and hole doping can inject carriers into the conducting layer that leads to the superconductivity [7,8]. Within a very short time, the transition temperature has been doubled from 26K to more than 50K. The microscopic theory of superconductivity proposed by Bardeen, Cooper, and Schrieffer (BCS theory) explains the superconductivity of certain metals, in which the paired electrons are regarded as operating in a single quantum state and can move freely without resistance [9]. The theory of superconductivity in the copper-based family of ceramic superconductors is still the basis of considerable research [10]. Superconductivity seems to be related to magnetic behavior in copper based superconductors, where magnetism is likely to be involved in high temperature superconductivity. The parent material of novel FeAs based superconductors is antiferromagnetic. As magnetism can exist over a very broad range of temperatures, the field is moving towards higher transition temperatures that can encompass the physiological range [11,12]. An important question relates to their toxicity [11]. Arsenic and some arsenic compounds are known to be severely cytotoxic in their free state, which raises the possibility that FeAs-based superconductors might be precluded from future biological studies, and consequently their presentation at a cellular interface. In this manuscript we present novel *in vitro* evidence that cells can be grown on the PrO$_x$FeAs (x=0.75) superconductor material for a prolonged period of time without any deleterious effects to the cells.

## 2. Experimental methods

*$PrO_xFeAs$ (x=0.75) superconductor*. The superconductor was prepared at 6 GPa and 1350 ºC by using PrAs, Fe and $Fe_2O_3$ as the starting materials. The nominal composition of the superconductor was $PrO_xFeAs$ (x=0.75). Electrical transport measurement showed a transition temperature Tc of about 45 K with zero resistance below 33 K. [Detailed information can be found in arXiv: 0806.2379] For cell based assays, slices with thicknesses of approximately 0.5 mm were taken from the superconductor pellet using a diamond saw, after which the surfaces of the slices were polished to optical smoothness. Superconductors were subsequently sterilized for cell culture purposes (vide infra) by 30 min exposure to ultraviolet light.

*Cell culture & microscopy analysis*. HeLa cells and non-immortalized human diploid dermal fibroblasts (GM3348) were cultured in Advanced Dulbecco's modified Eagle's medium (DMEM) with 5% fetal bovine serum (FBS) supplemented with 1% penicillin, streptomycin and amphotericin B (Invitrogen). The cells were incubated at 37ºC, 5% $CO_2$ with 100% relative humidity. For qualitative analysis, $10^4$ HeLa cells or $10^5$ dermal fibroblasts were seeded and cultured 50 mm tissue culture dishes (BD Biosciences) with superconductors placed at the center of each culture dish. Images were taken at the time of seeding, 10 hrs and 24 hrs after seeding using a light microscope (Olympus ULWCD 0.30) under transmitted imaging mode. The specimens were then fixed using 2.5% glutaraldehyde and 1% $OsO_4$, dehydrated up to 100% ethanol then further dried with HMDS for 3 minutes. Samples were coated with 10 nm gold film and subsequently viewed using a field emission scanning electron microscope (Zeiss Ultra) to study overall structural organization and behavior of cells grown on $PrO_xFeAs$ material.

*Viability assay*. A quantitative cell viability study was performed with cells cultured in the presence of superconductors and was carried out according to ISO10993-5 standard test protocol. Briefly, superconductor specimens were prepared and placed in wells of a 24-well culture plate. Next, $10^4$ dermal fibroblasts were seeded in each well that contained the superconductor material. Controls comprised of wells without the presence of $PrO_xFeAs$. CellTiter 96 Aqueous One Solution Reagent (Promega) was used to determine the total number of viable cells at time intervals 0 hr, 24 hr, and 72 hr after seeding.

3. Results

As shown in figure 1, light microscopy investigations show HeLa cells are able to grow well around the superconductor. Upon seeding (Figure 1a), HeLa cells were spherical and suspended in the medium. As early as ten hrs after seeding (Figure 1b), the majority of the cells adhered to the bottom of the culture dish and cells were elongated. Twenty-four hrs after seeding (Figure 1c), HeLa cells adopted a classical spread out, polygonal appearance and cell numbers appeared to increase substantially as a function of time as compared to 10 hrs post-seeding. Cells appeared to be actively dividing as indicated by the presence of rounded up mitotic cells. Similar results were obtained with optical microscope studies on dermal fibroblasts (data not shown).

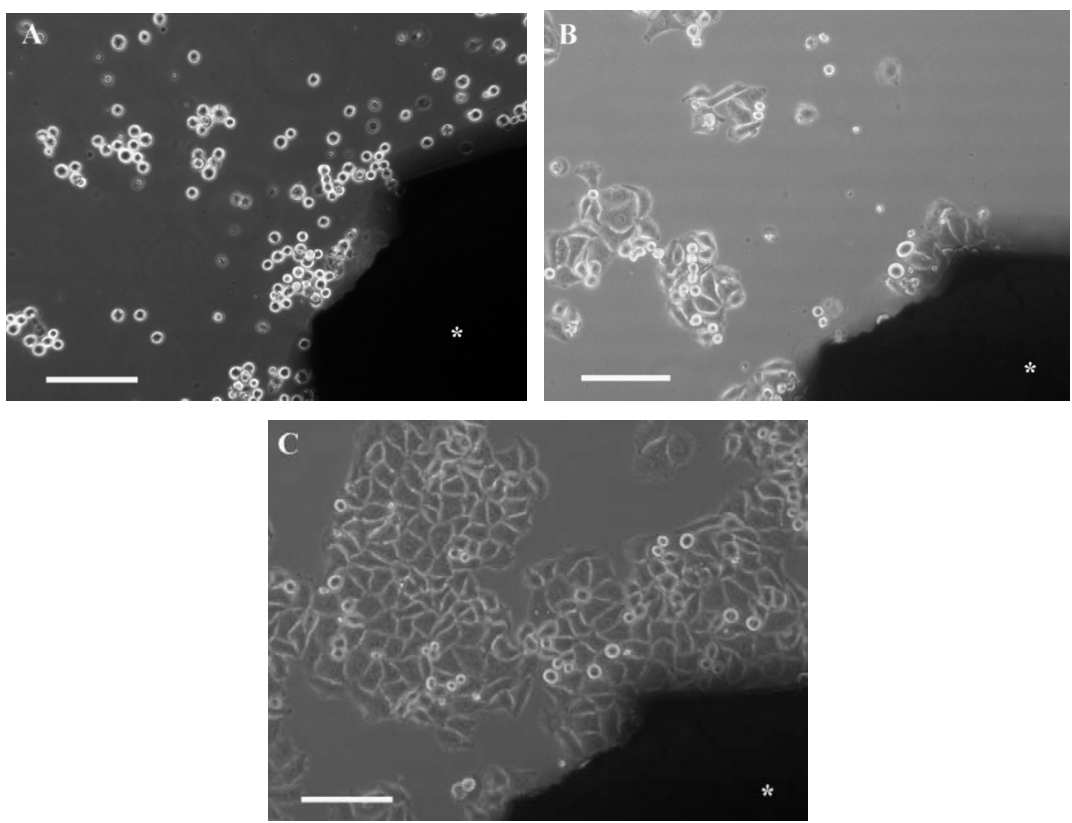

**Figure 1**. Transmission light optical microscopic images of HeLa cells near the superconductor (*) at different stages of growth. Note the apparent changes in cell shape and number near the superconductor indicate favorable conditions for growth. (**A**) 0 hour, (**B**) 10 hours and (**C**) 24 hours after seeding.

After 24 hrs, HeLa cells and skin fibroblasts were investigated by scanning electron microscopy. HeLa cells were dispersed uniformly on the surface of the superconductor. Along with a large number of polygonal-shaped HeLa cells, many of them were observed to be in the mitotic stage of the cell cycle as evidenced by the more spherical-like morphology depicted in Figure 2b. A HeLa cell undergoing cytokinesis can be seen in Figure 2c where only a thin fiber connects the two nascent daughter cells, indicating that cells grown on the superconductor actively divide. Furthermore, figure 2d shows a HeLa cell on the unpolished surface of the superconductor, demonstrating that the superconductor substrate does not have to be flat to allow cell attachment. Note the clearly visible laminated structure of the superconductor.

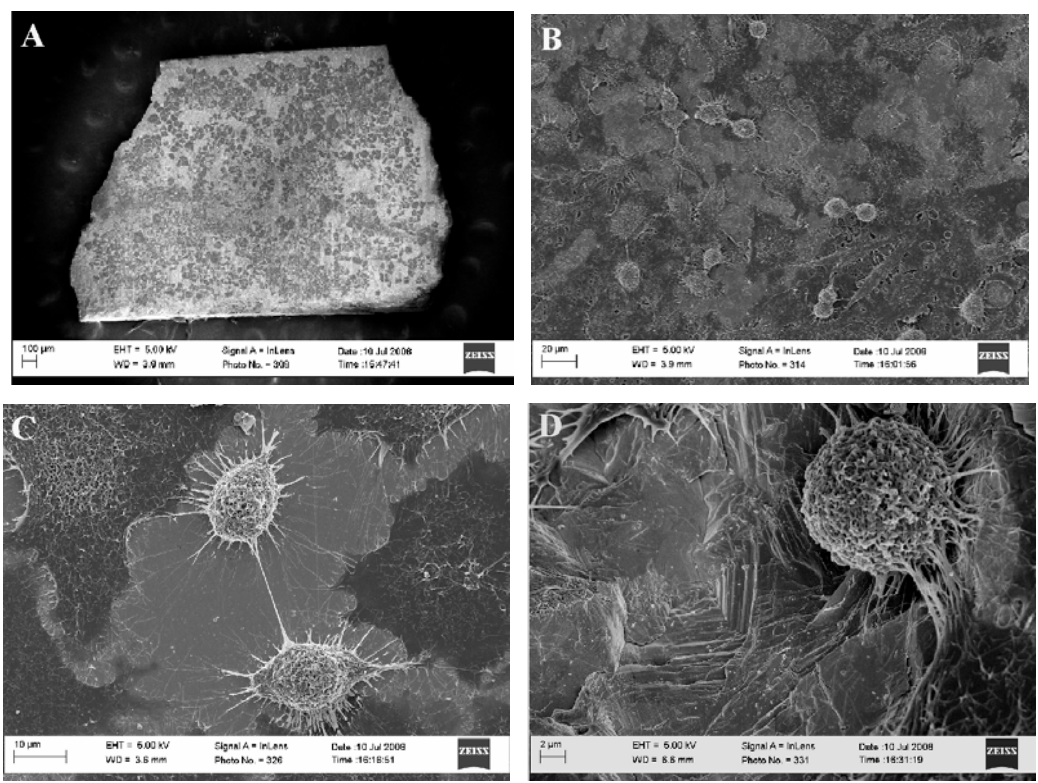

**Figure 2**. SEM images of HeLa cells grown on the superconductor for 24 hours. (**A**) Low magnification image showing cell adhesion on the superconductor surface. (**B**) High magnification image of cells in figure 2A. Rounded up cells are indicative of cells undergoing mitosis. (**C**) A cell undergoing cytokinesis. (**D**) An image of the side surface of the superconductor. The laminated superconductor and a mitotic cell can be clearly observed.

HeLa cells are an immortalized cell line derived from cervical cancer. We extended our studies to normal diploid skin fibroblasts and found that the cell adhesion behavior of dermal fibroblasts was similar to that observed in HeLa cells. After 24 hr growth, cells were prepared for SEM. Figure 3 shows representative SEM images of dermal fibroblasts. It can be concluded that, similar to the HeLa cells, the dermal fibroblasts can grow well around and on the surface of the superconductor.

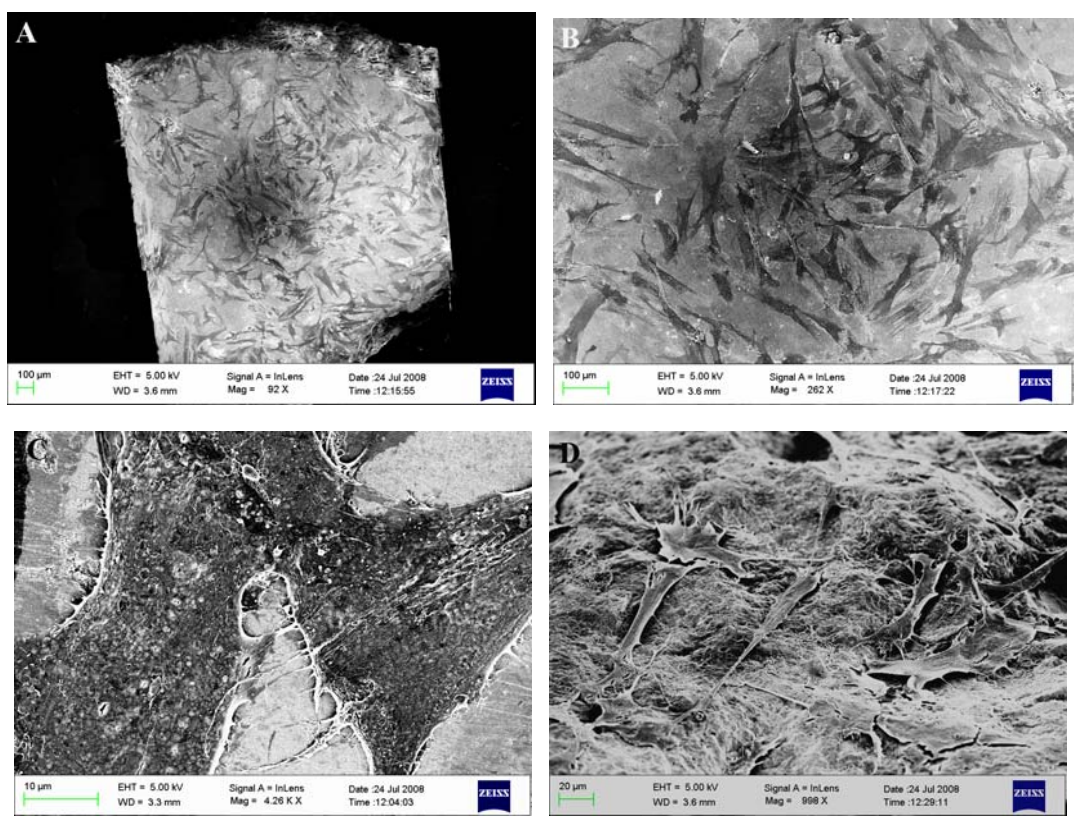

**Figure 3**. SEM images of dermal fibroblast cells grown on the superconductor surface for 24 hours. (**A**) Low magnification of the specimen with spindle shaped dermal fibroblast cells on the superconductor surface. (**B**) Intermediate magnification of the specimen in the same area showing the overall spindle-like shape of the fibroblasts. (**C**) Higher magnification imaging reveals that the cells are well attached on the surface of the superconductor. (**D**) Note, intermediate magnification image of dermal fibroblasts cells grown on the side surface of the superconductor.

We next determined the cell viability. The absorbance values for control and cells cultured with superconductor are plotted in Figure 4. It is apparent that the absorbance increased with the cell culturing duration for both control and superconductor. T-test analysis showed that cell viability and proliferation of the test group were the same as that of the control group with >90% confidence for all time periods.

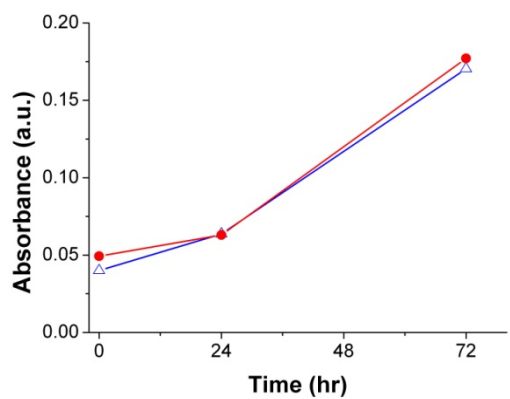

**Figure 4**. Absorbance reading for control (solid circles) and cells cultured (open triangles) with superconductor at time 0, 24 hours and 72 hours after seeding.

## 4. Discussion

Recent *in vitro* studies have demonstrated that the presence of arsenicals can induce cell death [13,14] by disrupting the citric acid cycle by allosteric inhibition of sulfhydryl metabolic enzymes or competition with phosphate in ATP synthesis leading to rapid hydrolysis of ATP. In each of these cases, the energy conversion processes of cells were adversely disrupted by arsenicals which finally resulted in cell death. Had the superconductor $PrO_xFeAs$ material described in this study been severely toxic to cells, HeLa cells and skin fibroblasts would have not been able to adhere and grow for a prolonged time. Furthermore, had the superconductor $PrO_xFeAs$ slowly leached out toxic particles, the cells at the periphery of the superconductor specimen would have been structurally different to cells grown away from the $PrO_xFeAs$ material. In this study, we successfully demonstrated that cells in the presence of $PrO_xFeAs$ specimen were able to adhere and proliferate, providing support for the biocompatibility aspects of $PrO_xFeAs$-based superconductor material. The proliferation of HeLa cells was particularly visible by the presence of adhered cells 10 hr and 24 hr after seeding. This data suggests that the superconductor $PrO_xFeAs$ is not evidently toxic and does not leach out detectable levels of toxic particles into the culture medium.

We employed HeLa cells in this study as they grow rapidly and as such cell proliferation is easily determined. Another useful feature of HeLa cells is that they change from polygonal flat patch shapes to spherical cells upon the initiation of mitosis. After cell division is complete, the two daughter HeLa cells will convert back to polygonal flat patches. This morphological feature makes mitosis easy to observe under optical microscopy and scanning electron microscopy. As HeLa cells are a cell line derived from cervical cancer cells, their cellular responses to superconductor might be different from primary cultured cells. Therefore, normal human dermal fibroblasts were also analyzed. We demonstrated that these cells were also able to adhere and grow on top of the superconductor specimens. These data demonstrate that the superconductor $PrO_xFeAs$ is most likely not detectably cytotoxic for cells as assessed by microscopic means.

To confirm our microscopy findings we performed quantitative cell viability studies. Human dermal fibroblasts were used to evaluate the possible adverse effects of the superconductor on cell viability over a 4 day period. The cell viability assay adopted in this paper is a colorimetric method for determining the number of viable cells in proliferation or cytotoxicity assays. The MTT tetrazolium compound present in the reagent is bioreduced by cells into a colored formazan product that is soluble in tissue culture medium and absorbs light of 490nm wavelength. A higher absorbance represents a higher viability and proliferation of cells. From day 1 to day 3, there was a steady increase in absorbance indicating increased cell proliferation. There was no statistical difference between the viability of control cells and cells cultured with superconductors, indicating the superconductor has no detectable detrimental effect on cells.

## 5. Conclusion

Cells remained viable and proliferative when cultured in direct contact with the superconductor $PrO_xFeAs$ (x=0.75).

Acknowledgments

This work was financially supported by the NSFC (10774068), NCET(07-0430) and 973 Program (2006CB921800). The authors acknowledge the facilities as well as technical assistance from staff at the Australian Key Centre for Microscopy and Microanalysis (AKCMM) of The University of Sydney. The first author also acknowledges the financial support from the University of Sydney under the "*International Visiting Research Fellow*" scheme.